\documentclass[12pt,a4paper]{article}
\usepackage[utf8]{inputenc}
\usepackage[T1]{fontenc}
\usepackage{amsmath}
\usepackage{amssymb}
\usepackage{tikz}
\usepackage{tabularx}
\usepackage{authblk}
\usepackage[title]{appendix}
\usepackage{graphicx}
\usepackage{color} 
\usepackage[linktoc=page]{hyperref}
\hypersetup{
	colorlinks=true,
	linkcolor=blue,
	urlcolor=blue,
	bookmarks=true,
	citecolor=[rgb]{0.54,0,0}
}
\author{Yoav Zigdon}
\affil{{\normalsize \textit{School of Physics and Astronomy, Tel Aviv University, Ramat Aviv, 69978, Israel}} \\
	{\normalsize yoavzi(at)tauex.tau.ac.il}}
\date{}
\title{de Sitter in String Theory vs. Gibbons \& Hawking}

\begin{document}
	\maketitle
	\begin{abstract}
		This paper corroborates a statement that perturbative string theory does not admit a solution whose spacetime metric is de Sitter times a closed manifold, to all orders in the $\alpha'$ and $g_s$ expansions, under the assumption that the logarithm of the sphere partition function of Euclidean quantum gravity receives a nonzero contribution proportional to $\frac{1}{G_N}$ in a saddle-point approximation. This assumption is related to the Gibbons-Hawking proposal that the entropy of the cosmological horizon of the static patch is $\frac{A}{4G_N}$. Evidence for the statement comes from independent approaches to the effective action of string theory, all of which agree that the tree-level action vanishes for closed Euclidean target-space solutions. One possible implication is that the state of the Universe will depart from an asymptotically de Sitter spacetime.  
	\end{abstract}
	\newpage
	\tableofcontents
	\section{Introduction} 
	In 1998, a report on observations of type Ia supernovae provided strong evidence that the expansion of the Universe is accelerating \cite{SupernovaSearchTeam:1998fmf}. An asymptotically de Sitter spacetime provides a natural theoretical description of such late-time acceleration, capturing the large-scale dynamics of the Universe in multiple cosmological models. However, recent data from the Dark Energy Spectroscopic Instrument (DESI) \cite{DESI:2025zgx} favor scenarios in which the acceleration decreases with time, challenging a strictly de Sitter future. The objective of this paper is to support this phenomenological indication, from an independent theoretical perspective, that the asymptotic future of the Universe deviates from de Sitter. 
	
	In M-theory and string theory, two general approaches have been developed in relation to the de Sitter model of the accelerated expansion of the Universe. In the first, ``positive approach'', one aims at constructing specific solutions to the low-energy field equations that contain a warped de Sitter part of the spacetime manifold, utilizing extra dimensions, quantized fluxes, and extended sources. In the second, ``negative approach'', the goal is to prove or argue that certain assumptions imply the absence of such warped solutions. 
	
	Examples of papers that adopted the ``positive approach'' include~\cite{Maloney}-\cite{ValeixoBento:2025yhz}. These papers wrote potentials for moduli (or a single modulus) in effective field theories and found either a maximum or a minimum of positive energy density. One of the aspects of these papers is the absence of parametrically small parameters that would justify keeping a subset of terms in the potentials while dropping the rest of the contributions,  which appear in the exact potentials of the full string/M-theory. \footnote{Coupling parameters in the Standard Model are also not ``parametrically small'', and nonetheless perturbation theory provides accurate predictions for experimental outcomes in particle accelerators, at least in the electroweak sector.} Moreover, some of the terms in the potential are unknown to date. Reference~\cite{Sethi:2017phn} emphasized the lack of truncations of M-theory and string theory down to supergravity and a selected set of higher-derivative terms.   While there is much more to be said about the proposed realizations of de Sitter in string/M-theory, the present paper neither investigates any of them nor does it propose a new construction.
	
	The ``negative approach'' includes a set of no-go theorems and arguments pioneered by Gibbons~\cite{Gibbons:1984kp},\cite{Gibbons:2003gb}, who showed that, in the context of supergravity, the strong energy condition, or the requirement that the time-time component of the Ricci tensor is non-negative, and the presence of extra compact dimensions forming a closed manifold, which is associated with a smooth and non-vanishing warp factor, rule out warped de Sitter compactifications. See also  \cite{deWit:1986mwo}.
	
	In 1985, reference~\cite{Dine:1985he} listed three possible behaviors of the asymptotic potential for the dilaton:  identically zero,  positive, or negative, and concluded that in likely scenarios, a minimum appears when string theory is strongly coupled. This would exclude de Sitter vacua in string theory when perturbation theory applies. 
	 
	One of the no-go theorems against de Sitter in reference~\cite{Maldacena:2000mw} assumes a non-positive potential for scalar fields within  supergravity and a finite Newton constant when reducing on extra compact dimensions, combined with certain assumptions about the warp factor.  Reference~\cite{Hertzberg:2007wc} included fluxes, D6-branes, and O6-branes in Type IIA superstring theory compactified on a Calabi-Yau threefold at weak string coupling and low curvatures to show that de Sitter vacua do not exist.  An extension to systems with both D$q$ and O$q$ branes (in Type IIB superstring theory for odd $q$), and  possibly curved compact manifolds, is written in~\cite{Obied:2018sgi}. Methodologies in the proofs of no-go theorems in supergravity typically involve integrating one of the equations of motion, or a linear combination of equations (times a power of the warp factor), over the compact internal dimensions.~\footnote{Many more references worked in the framework of supergravity to make no-go arguments; see~\cite{Andriot:2026lac} and references therein.}
	
	In the context of weakly coupled heterotic string theory, including leading inverse tension, or $\alpha'$ corrections to produce a no-go argument is due to~\cite{Green:2011cn}, which was generalized to all orders in $\alpha'$, assuming also that tensor potentials  on the four-dimensional part of the would-be warped de Sitter are pure gauge~\cite{Gautason:2012tb}. 
	
	A no-go theorem that applies in the strong curvature regime has appeared in reference~\cite{Kutasov:2015eba} that examined whether Euclidean continuations of (an unwarped) de Sitter spacetimes in dimensions $d\geq 4$  exist as target spaces of worldsheet conformal field theories, with a heterotic super-algebra structure. It was found that the principles of unitarity and the Kac-Moody algebra of the corresponding $SO(d+1)_k$ Wess-Zumino-Witten (WZW) models rule out $8\leq d$. For $4\leq d\leq 7$, discrete sets of candidate (bosonic) levels of the Kac-Moody algebra were found to be of order one, meaning that  parametrically large de Sitter spacetimes are inconsistent.
	  
	The outline of this paper is as follows. In Section~\ref{sec:theorem}, a claim is phrased whose conclusion is the absence of any solutions of the form of products of closed compact manifolds and de Sitter in perturbative string theory. This new no-go result is argued to be valid to all order in $\alpha'$ and the string coupling $g_s$, assuming that a version of the Gibbons-Hawking proposal~\cite{GH} is correct. Section~\ref{sec:proof} contains evidence for the claim produced from different approaches to the target space effective action of string theory. Section~\ref{sec:disc} includes a discussion about the assumption of the argument and its conclusion. Appendix~\ref{app:sug} shows that the supergravity actions that arise from Type IIB superstring theory and M-theory are boundary terms on shell. Appendix~\ref{app:Ieff} provides an expression for the quantum effective action of perturbative string theory on shell.  
	\section{Claim}
	\label{sec:theorem}
	Suppose that the Euclidean quantum gravity path integral $Z$ is well-defined and has a reliable saddle-point approximation coming from a sphere solution, and that $\log(Z)$ receives a contribution proportional to $\frac{1}{G_N ^{(D)}}$ from that saddle, where $G_N ^{(D)}$ is the $D$-dimensional Newton constant.
	 	
	Then perturbative string theory, treated as a theory of quantum gravity, does not contain solutions in which the target spacetime is a direct product of a non-singular compact manifold without boundaries and a $d$-dimensional de Sitter spacetime. This holds to all orders in perturbation theory in the inverse string tension $\alpha'$ and the string coupling $g_s$ (a relevant definition of ``$g_s$'' is written in subsection~\ref{sub:heart}).
	
	In addition to the assumption above, if a warp factor of a non-singular, compact, boundary-free internal manifold is smooth and finite, then a warped de Sitter compactification does not exist within perturbative string theory. 
	
	As reviewed in subsection~\ref{subs:GH}, the original Gibbons-Hawking proposal \cite{GH} was phrased in the context of semiclassical gravity with a positive cosmological constant. The assumption above departs from that original setup, and can be thought of as a version of the Gibbons-Hawking proposal for the setup of perturbative string theory. Adding a thermodynamic interpretation to the assumption, one obtains that in quantum gravity, the entropy of the cosmological horizon still has a nonzero piece proportional to $\frac{1}{G_N ^{(D)}}$.
	\section{Evidence}
	\label{sec:proof}
	\subsection{The on-shell action of a compact solution vanishes}
	\label{subs:onshell}	
	The objective of this subsection is to review that several approaches and methods lead to the conclusion that the effective, on-shell, tree-level action of string theory vanishes on any smooth closed target space solution. Several general comments follow.
	
		The effective action of string theory ``$I$'' is generally an off-shell functional, constrained by the requirements that (a) the string S-matrix in flat spacetime is equal to the S-matrix derived from $I$, and (b) the conditions of Weyl invariance of the worldsheet theory for string propagation in a target spacetime $M$ are equivalent to the equations of motion derived from $I$.
		
			It is important to note that the object $I$ is ambiguous, as one could, for instance, integrate terms by parts and drop the surface term. Another source of ambiguity is that for any field in the effective action $T$ with mass $m\neq0$, $-\frac{1}{m^2}\nabla^2 T$ is indistinguishable from $T$. However, the ambiguities will not affect the conclusion below. 
			
				Additionally, at least in principle, $I$ could be appended by boundary terms $I_{\text{bdy}}$ that make the variational principle well defined. The main result of the next sub-subsection is independent of such terms.
				
				Localized sources such as NS5-branes, D-branes, O-planes, and fundamental strings will not be included in the effective tree-level action because they generate classical singularities in target spacetime. (Although, from a worldsheet perspective, known separated NS5-branes solutions are free of any singularities~\cite{Giveon:1999px},\cite{Giveon:1999tq},\cite{Martinec:2017ztd}.) 
	\subsubsection{The on-shell action is a boundary term}
	The main technical tool in one approach is the expression of the on-shell, tree-level part of $I$ as a boundary term. A class of theories has the property that their on-shell actions are purely surface terms: Pure Maxwell theory, pure General Relativity without a cosmological constant, and, as shown in Appendix~\ref{app:sug}, 10-dimensional Type IIB supergravity in the democratic formalism and 11-dimensional supergravity. This property follows from the existence of a classical scaling symmetry that multiplies the action by a constant while preserving the equations of motion. More examples, including supersymmetric Yang-Mills theories, appear in reference~\cite{Jevicki:1998ub}.~\footnote{I thank Aron Wall for emphasizing that a variety of theories exhibit a classical scaling symmetry and its implications.}  
	 
	The string-frame effective actions of Type II and the heterotic string theories at tree-level in $D$-dimensions can be formally expressed as:
	\begin{equation}
		\label{I}
		 I_{\text{tree-level}}= \int_M d^{D} x ~e^{-2\Phi} L_{-2}~~,
	\end{equation}
	where $M$ denotes the target spacetime, $x$ is a collective notation for $D$ coordinates, $\Phi$ is the dilaton, and $L_{-2}$ is a sum of functions built from target spacetime fields which do not depend on $\Phi$ itself, but generally do depend on derivatives of $\Phi$. The object $e^{-2\Phi}L_{-2}$ contains the supergravity Lagrangian density and $\alpha'$ corrections thereof. 	\footnote{A comment is that the Dine-Seiberg runaway problem~\cite{Dine:1985he} is already reflected in Eq.~(\ref{I}) when fixing all fields to constant values such that $L=\text{constant}$. In that case, the potential for the dilaton is either identically zero, positive, or negative - thereby excluding a de Sitter solution at parametrically weak coupling. The present discussion is more general, in that generic classical solutions display fields that vary in spacetime (in particular, the dilaton), rather than being frozen at constant values. Also, perturbatively small corrections in $g_s$ will be incorporated in subsection~\ref{sub:heart}.}
	
	In Type II superstring theory, a different convention for Ramond-Ramond (RR) field strengths $F_p'$ is adopted. The $F_p'$ field strengths are related to the ``conventional'' RR field strengths $F_p$ whose integrals on compact manifolds are quantized, via  
	\begin{equation}
		\label{Fpp} 
		F_{p}'= e^{\Phi} F_p~~,
	\end{equation}
	as in Eqs.~(12.1.12) of the second Polchinski string theory textbook \cite{Polchinski:1998rr}. The field strength forms $F_p '$ depend on wedge products of the RR potentials $C_{p-1} '$ and the derivative of the dilaton.
	(One need not adopt this convention in Type II supergravity in the democratic formalism -  Appendix~\ref{app:sug} shows that its on-shell action is still a boundary term.)  
	
	 The dimension $D$ is not necessarily set to $10$, allowing for a nonzero cosmological constant at tree-level
	\begin{equation}
		\label{CC}
		\Lambda = \frac{D-10}{3\alpha'}~.
	\end{equation}
	Relative to the Ricci scalar $R$, the cosmological constant in Eq.~(\ref{CC}) appears in the effective Lagrangian density as $e^{-2\Phi}(R-2\Lambda)$.
       
	Now, it is a known result due to \cite{Callan:1986jb},\cite{CMW} that the on-shell action of any solution of the classical string theory equations of motion is a boundary term (see also \cite{KT},\cite{BZ}). Here is a review of this fact: The dilaton equation of motion reads
	\begin{align}
		\label{dil} 
		-2 e^{-2\Phi} L_{-2}=\partial_{\mu} \Big(e^{-2\Phi}\frac{\delta L_{-2}}{\delta (\partial_{\mu} \Phi)} \Big)-\partial_{\mu} \partial_{\nu}\Big(e^{-2\Phi}\frac{\delta L_{-2}}{\delta (\partial_{\mu} \partial_{\nu} \Phi)} \Big)+... ~~.
	\end{align}
	Each term on the R.H.S. of Eq.~(\ref{dil}) that has a variation of $L_{-2}$ with respect to $k$ derivatives of $\Phi$, holds fixed $n\neq k$ derivatives of $\Phi$. The dots ``...''  refer to possible dependence of $L_{-2}$ on three or more derivatives of $\Phi$. Multiplying Eq.~(\ref{dil}) by $-\frac{1}{2}$ and then substituting $ e^{-2\Phi}L_{-2}$ into the action in Eq.~(\ref{I}), one obtains that the effective, on-shell, tree-level action is given by
	\begin{align}
		\label{Icl}
		I_{\text{tree-level,on-shell}} &= -\frac{1}{2} \int_M d^{D} x ~\partial_{\mu} \Bigg(e^{-2\Phi}\frac{\delta L_{-2}}{\delta (\partial_{\mu} \Phi)}-\partial_{\nu}\Big(e^{-2\Phi}\frac{\delta L_{-2}}{\delta (\partial_{\mu} \partial_{\nu} \Phi)}\Big)+...\Bigg)~~.
	\end{align}
	In this equation, the on-shell solution has been substituted on the R.H.S.\\ 
	By the generalized Stokes' theorem, the total derivative term in Eq.~(\ref{Icl}) is a boundary term with $n_{\mu}$ denoting a normal vector to the boundary: 
		\begin{align}
			\label{bdy}
		I_{\text{tree-level,one-shell}} &= -\frac{1}{2} \int_{\partial M} d^{D-1} x ~n_{\mu} \Bigg(e^{-2\Phi}\frac{\delta L_{-2}}{\delta (\partial_{\mu} \Phi)}-\partial_{\nu}\Big(e^{-2\Phi}\frac{\delta L_{-2}}{\delta (\partial_{\mu} \partial_{\nu} \Phi)}\Big)+...\Bigg)~~.
	\end{align}
    In applying the generalized Stokes' theorem, the smoothness and well-defined nature of the expression in the parentheses of the integrand on the R.H.S. of Eq.~(\ref{Icl}) are assumed.  
    
	Therefore, the action $I_{\text{tree-level,on-shell}}$ is a  boundary term that depends only on the asymptotic falloff behavior of target-space fields near the boundary $\partial M$. In the special case where $M$ is a closed Euclidean space, the surface terms drop, and it follows that
	\begin{equation}
		\label{I0}
		I_{\text{tree-level,on-shell}}=0~~.
	\end{equation}
	    	    An important qualification is discussed\footnote{I thank Aron Wall for writing this example.}. It is not true in general that if the on-shell action $I_{\text{cl}}$ of a given quantum field theory defined on a closed manifold is a surface term, then $I_{\text{cl}}=0$. For example, in a two-dimensional pure Maxwell theory on the sphere, the action is expressible in terms of a boundary term. However, any classical solution in this theory with a nonzero constant field configuration has a nonzero on-shell action. This can be understood as a consequence of the gauge field not being globally well-defined on the entire manifold, so the boundary term is not gauge-invariant. Thus, one could wonder whether string theory, whose spectrum contains multiple gauge fields, also falls into this category of theories where, despite the classical action being a boundary term, it does not vanish on a closed surface. Some evidence exists that this is not the case: the following two sub-subsections, which rely on different approaches to closed bosonic string field theory (rather than Type II or the heterotic string theories), reach the same conclusion.  Also, in the special case where $M$ is Euclidean and asymptotically flat, and RR fields are turned off, \cite{CMW} simplified $\text{I}_{\text{tree-level,on-shell}}$. In that case, Eq.~(\ref{bdy}) receives a contribution only from the kinetic term of the dilaton, giving 
	    \begin{equation}
	    	\label{Ibdy}
	    	I_{\text{tree-level,on-shell,asym-flat}} =-\frac{1}{8\pi G_N ^{(D)}}\int_{\partial M} d^{D-1}x \sqrt{h} n^{\mu} \partial_{\mu} \big(e^{-2\Phi} \big)~.
	    \end{equation} 
	    See Appendix~\ref{app:sug} in this paper for a rederivation of this result in Type IIB supergravity in the democratic formalism. In particular, for the bosonic and heterotic string theories in asymptotically flat space, one could apply Eq.~(\ref{Ibdy}), which is independent of any gauge fields (such as the NS-NS two-form and the gauge fields of the heterotic string). In addition, reference~\cite{Hawking:1995ap} showed that supplementing the action of the (Euclidean) Einstein-Maxwell theory with a boundary term that depends on the gauge potential has the effect of describing states in a fixed electric charged ensemble. In contrast, no such boundary term is needed to work in a fixed magnetic charge ensemble. Similar boundary terms are relevant in Type IIB in the democratic formalism and 11-dimensional supergravities, as shown in appendix~\ref{app:sug}. One could choose to commit to a particular ensemble (either grand-canonical or a superselection sector with fixed charges) where such boundary terms are absent - thereby removing the apparent non-gauge invariance of the action on a closed target. For these reasons, the qualification described here is considered invalid for tree-level string theory. The conclusion is that the tree-level effective action of string theory vanishes on a closed solution. 
		\subsubsection{Tseytlin's prescriptions}
	References~\cite{Tseytlin:1987bz},\cite{Tseytlin:1988tv},\cite{Tseytlin:1988rr} proposed and corroborated a prescription that expresses the tree-level effective action of the massless sector of closed bosonic string theory in terms of a derivative of the renormalized sphere partition function $Z_{\mathbb{S}^2}$ of the worldsheet theory relative to $\log(\mu)$, where $\mu$ is a renormalization scale:  
	\begin{equation}
		\label{Ibos}
		I_{\text{tree-level}} = \frac{\partial Z_{\mathbb{S}^2}}{\partial (\log(\mu))}~.
	\end{equation}
	The variation in Eq.~(\ref{Ibos}) subtracts off a logarithmically-divergent volume of the $\text{SL}(2,\mathbb{C})$ group for a spherical worldsheet. Reference~\cite{Tseytlin:2000mt} proposed an improved prescription that makes the tachyon tadpole vanish in the bosonic string (whereas in Type 0 string theory the previous prescription suffices). \footnote{References \cite{Ahmadain:2022tew},\cite{Ahmadain:2024hdp} further developed this formalism.}
	The papers have argued that there exists a scheme whereby the tree-level effective action is given by
	\begin{equation}
		\label{S1}
		I_{\text{tree-level}} \propto \int d^D x \sqrt{G} e^{-2\Phi} \widetilde{\beta}^{\Phi}~.
	\end{equation} 
	The quantity $\widetilde{\beta}^{\Phi}$ is referred to as the ``central charge coefficient'' and for the massless sector of the bosonic string is linear in two of the $\beta$-functions of the dilaton and metric couplings. 
	Moreover, this quantity appears in the expectation value of the trace of the stress-energy tensor of the worldsheet theory,
	\begin{equation}
		\langle T^a _a \rangle = \frac{1}{4\pi}\widetilde{\beta}^{\Phi} R^{(2)}+\text{non-local terms}~,
	\end{equation}
	where $R^{(2)}$ is the Ricci scalar on the worldsheet. Therefore, the condition that the worldsheet theory is conformal implies that $\widetilde{\beta}^{\Phi}=0$, meaning that the on-shell Lagrangian density in Eq.~(\ref{S1}) and its integral vanish: 
		\begin{equation}
			\label{S2}
			\widetilde{\beta}^{\Phi}=0\Rightarrow I_{\text{tree-level}} =0~.
		\end{equation} 
	 This is the main result of this sub-subsection. Note that $\widetilde{\beta}^{\Phi}=0$ is stronger than a statement that the \underline{integrated} Lagrangian density vanishes on a closed target. 
	\subsubsection{A bosonic string field theory argument}
	In the third approach, namely closed bosonic string field theory, a different calculation valid for a compact target space has appeared in reference~\cite{Erler}. Given $Q$, the nilpotent BRST operator, $\Psi$, the ghost-two closed string field, $\kappa$, a coupling constant proportional to the closed string coupling, $\omega$, the symplectic form, and $L_n(...)$ denoting Grassmann-odd string products, the starting point is a tree-level closed, bosonic string field theory action given by
	\begin{equation}
		I_{\text{SFT}} = \frac{1}{2} \omega \big(\Psi,Q \Psi\big)+\sum_{n=3} ^{\infty} \frac{\kappa^{n-2}}{n!}  \omega\big(\Psi,L_{n-1} \big(\Psi^{n-1}\big)\big)~.
	\end{equation}
	A result in~\cite{DilatonTheorem} was used, where a field variation ``U'' that rescales the coupling constant $\kappa$ was shown to exist:
	\begin{equation}
		\delta_U \Psi = \kappa \frac{d}{d\kappa} I_{SFT}~.
	\end{equation}
	Using this variation of the string field, (a) the variation of the closed string field theory action is zero, and (b) that variation produces the action time a constant different than one, implying that the on-shell action vanishes:
	\begin{equation}
		I_{\text{SFT,on-shell}}=0~.
	\end{equation}
	\subsubsection{Intermediate conclusion}	
	The inclusion of perturbative corrections in the string coupling in an effective action approach modifies the result in Eq.~(\ref{I0}). In Appendix~\ref{app:Ieff}, it is shown that given the general effective action (of the same form as Eq.~(27) of reference~\cite{Tseytlin:1988tv})
	\begin{equation}
		\label{I2}
		I= \int_M d^{D} x ~\Big(e^{-2\Phi} L_{-2}+\sum_{g=0} ^{\infty} e^{g\Phi} L_g\Big)~~,
	\end{equation}
	an on-shell evaluation of $I$ on a closed manifold (using the dilaton equation and the generalized Stokes' theorem) yields
	\begin{equation}
		\label{I02}
		I_{\text{on-shell}}=\int_M d^D x\sum_{g=0} ^{\infty} \Big(1+\frac{1}{2}g\Big)e^{g\Phi} L_{g}~.
	\end{equation}
	The main conclusion of this subsection is that the on-shell effective action of perturbative string theory has no $\frac{1}{G_N}$ piece. This will be important in what follows.
	\subsection{Review of the Gibbons-Hawking proposal}
	\label{subs:GH}
	Next, a review of the Gibbons-Hawking proposal is written. Readers familiar with this material can skip to subsection~\ref{sub:heart}, although even some experts could learn from later parts of the present subsection. 
	
	In $d$-dimensional pure gravity with a positive cosmological constant $\Lambda>0$,  reference \cite{GH} showed that the on-shell action of the Euclidean $d$-sphere solution is nonzero and is proportional to the area of the sphere:
	\begin{equation}
		\label{ISd}
		I_{\mathbb{S}^d} = -\frac{\omega_{d-2}H^{-(d-2)}}{4G_N ^{(d)}}~.
	\end{equation}
	Here, $\omega_{d-2}$ is the area of a unit radius $(d-2)$-sphere\footnote{The authors set $d=4$ but the result (\ref{ISd}) holds for any $d>2$. Also, Eq.~(\ref{ISd}) can be extended to higher, D-dimensional theories of gravity with a $d$-dimensional sphere part of the geometry. In this case, one should interpret $G_N ^{(d)}$ as the Newton constant in the $d$-dimensional theory obtained by a dimensional reduction from $D$ dimensions. }.

	Reference~\cite{GH} assumed that the sphere partition function $Z_{\mathbb{S}^d}$ has a reliable saddle-point approximation in which the first term of the logarithm of $Z_{\mathbb{S}^d}$ is related to the on-shell action:
	\begin{equation}
		\label{GH}
		\log(Z_{\mathbb{S}^d}) = -I_{\mathbb{S}^d}+\log(Z_{1-\text{loop}})+...~~.
	\end{equation}
	The one-loop determinant $Z_{1-\text{loop}}$ and the dots ``...'' in Eq.~(\ref{GH}) correspond to quantum corrections to the leading-order answer $-I_{\mathbb{S}^d}$, which contain a perturbative series in the parameter $G_N ^{(d)} H^{d-2}$ and also a logarithmic term proportional to $\log(G_N ^{(d)} H^{d-2})$ \cite{Anninos:2020hfj},\cite{Anninos:2021ihe}.  
	Further, Gibbons and Hawking interpreted the expression $\log(Z_{\mathbb{S}^d})$ as encoding the free energy of the cosmological horizon of the static patch of de Sitter spacetime. 
	Since the Arnowitt–Deser–Misner (ADM) energy is a boundary term, it vanishes on a closed manifold, and thus the logical implication of the Gibbons-Hawking proposal is that the tree-level term in the entropy associated with the cosmological horizon in the static patch background is 
	\begin{equation}
		\label{S}
		S_{\text{tree-level}}=-I_{\mathbb{S}^d}=\frac{\omega_{d-2}H^{-(d-2)}}{4G_N ^{(d)}}~~.
	\end{equation}
	A few comments are in order. First, the paper~\cite{GH2} interpreted this entropy as quantifying the lack of knowledge of an observer propagating in de Sitter spacetime about the state of the region beyond the cosmological horizon. Second, the first historical state-counting interpretation of this entropy was given in~\cite{MS} for $d=3$ in terms of highly excited states in a  chiral WZW model at an imaginary level. Third, in higher-derivative theories of gravity that admit a de Sitter background, references~\cite{Shu:2008yd},\cite{Gong:2023dwv} used the entropy function formalism due to~\cite{Sen:2005wa} in order to calculate corrections to the Gibbons-Hawking entropy. (Typically, that formalism is applied to calculate the entropy of extremal black holes in such theories.) Reference~\cite{Bobev:2022lcc} utilized the conventional Wald entropy formula instead for the same general purpose. Finally, section~\ref{sec:disc} contains a discussion about the validity of the assumption of~\ref{sec:theorem}, that a nonzero tree-level contribution to the Euclidean quantum gravity path integral arises, generalizing Eq.~(\ref{ISd}) to a regime with $\alpha'$ corrections.
	\subsection{Heart of the argument}
	\label{sub:heart}
	By way of contradiction, suppose a direct product of pure $d$-dimensional de Sitter and a $(D-d)$-dimensional closed manifold were to exist in perturbative string theory.   
	The target spacetime metric of the solution can be written in global coordinates as
	\begin{equation}
		\label{dS}
		ds^2 = \frac{1}{H^2} \Big( -dt^2 + \cosh^2 (t)d\Omega_{d-1}^2 \Big) + ds_{\mathcal{M}}^2~.
	\end{equation}
	Here, $H$ denotes the Hubble parameter, which is the inverse of the de Sitter length scale. $\mathcal{M}$ denotes a $(D-d)$-dimensional closed, compact and smooth manifold. In addition to this smooth metric, it is assumed that all other fields in the solution are smooth.  
	The solution (\ref{dS}) admits a Euclidean analytical continuation with an imaginary time $\tau=it$, such that the sigma model target space contains a $d$-dimensional sphere of radius $\frac{1}{H}$:
	\begin{equation}
		\label{EdS}
		ds^2 = \frac{1}{H^2} \Big( d\tau^2 + \cos^2 (\tau)d\Omega_{d-1}^2 \Big) + ds_{\mathcal{M}}^2~.
	\end{equation}
	This analytical continuation could generate complex NS-NS and RR potentials, but remains a solution of perturbative string theory.
	
	A gedanken experiment is discussed below to define ``$g_s$'' and ``$G_N ^{(D)}$'' in the background in Eq.~(\ref{EdS}). Typically, in asymptotically $\mathbb{R}^d \times \mathcal{M}_{10-d}$ space, one defines the Newton constant via the value of the exponential of the dilaton at infinity, which gives the asymptotic closed string coupling $g_s$. In Type II superstring theory, the 10-dimensional Newton constant $G_N ^{(10)}$ is related to $g_s$ via~\cite{Polchinski:1998rr}
	\begin{equation}
		\label{GN}
		G_N ^{(10)} = 8\pi^6 g_s ^2 (\alpha')^4~. 
	\end{equation}
	However, such a definition is not easily extendable to the sphere times internal manifold~(\ref{EdS}). The main reason is the absence of a well-defined S-matrix on a closed Universe, which would have allowed one to formally define the three-point interaction vertex of closed, short, weakly interacting strings and extract $g_s$. Lorentzian string amplitudes are generally defined by analytical continuations of vertex operators in the Euclidean target space, because in this way the string worldsheet path integral can converge. In the computation process of the Lorentzian string amplitudes, labels of the vertex operators are continued from the Euclidean target to the Lorentzian target~\cite{Jafferis}.
	
	 Practically, however, when the sphere is sufficiently large, $\sqrt{\alpha'} H\ll 1$, nearby observers who live in de Sitter could arrange for a scattering experiment of short strings to occur near the center of space and measure cross sections \underline{locally}, as done in particle accelerators on Earth, despite the possibility that we live in a closed cosmology. 
	 
	 Below, Eq.~(\ref{GN}) will be adopted for the assumed-to-exist background in Eq.~(\ref{EdS}), with the understanding that $g_s$ is defined as the expectation value of the exponential of the dilaton at the fixed point with respect to rotation transformations: $g_s = e^{\Phi(0)}$ (which should be further averaged over the internal manifold $\mathcal{M}$ when $\Phi$ has a non-trivial profile on $\mathcal{M}$). An alternative definition is to average the exponential of the dilaton over the entire space through $\bar{g}_s \equiv \frac{1}{\text{Vol}(M)}\int_M d^{D} x ~e^{\Phi}$ where $\text{Vol}(M)=\int_M d^D x$.
	   
	It is now explained that the perturbative string theory approach does not satisfy the version of the Gibbons-Hawking proposal phrased in assumption~\ref{sec:theorem}. A formal expression for the logarithm of the perturbative string theory partition function is
	\begin{equation}
		\label{Zresult}
		\log(Z_{\text{ST}})=-I_{\text{on-shell}}+\log(Z_{1-\text{loop}})+...~~, 
	\end{equation}
	where further perturbative quantum corrections are included in the dots. The first term on the R.H.S. of Eq.~(\ref{Zresult}) starts at order $O(g_s^0)$ because of the formula for the on-shell, quantum effective action in Eq.~(\ref{I02}). The second term similarly does not scale with $\frac{1}{G_N}$, and so are the terms in the dots, because they correspond to quantum corrections. The strict perturbative nature of the approach adopted here implies that the solution in question is not obtained by balancing classical effects with quantum effects, because the latter are perturbatively suppressed by powers of $g_s$ relative to the former. Formally, perturbative string theory requires a pre-existing classical background about which quantum corrections are calculated or expanded.  Thus, no $\frac{1}{G_N ^{(D)}}$ piece arises in the expansion~(\ref{Zresult}), in contrast to the assumed version of the Gibbons-Hawking proposal.
		
	Consequently, there is a contradiction between hypothesizing that a version of the Gibbons-Hawking proposal is correct, or more generally that there is a nonzero $\frac{1}{G_N ^{(D)}}$ contribution to the logarithm of the sphere partition function of quantum gravity, and the string theoretic result in Eq.~(\ref{I02}). One way to characterize this important difference between string theory and General Relativity is to note that the on-shell action of the latter, namely the Einstein-Hilbert action supplemented by a cosmological constant, is not a boundary term, leading to the fundamental contradiction noted here.

	Several immediate generalizations of the argument just given are written down. First, this contradiction also arises for small space-dependent deviations from pure de Sitter that are assumed to constitute solutions, if they preserve the global nature of its Euclidean continuation as a closed manifold. An effect of the ``small deviations'' is to perturb the area of the cosmological horizon of the static patch relative to the pure de Sitter value. Second, introducing warping into the ansatz (\ref{dS}) does not affect the result as long as the warp factor is finite and smooth so that the overall topology remains intact.   
	\section{Conclusion and discussion}
	\label{sec:disc}
	The conclusion is that either\\
	
	(a) de Sitter exists in perturbative string theory but fails to satisfy the version of the Gibbons-Hawking proposal where a nonzero tree-level piece is contained in the logarithm of the Euclidean path integral of string theory, \\
	
	or,\\
	
	(b) Perturbative string theory does not admit a de Sitter (times a closed manifold) background.\\ 
	
	The assumption of the argument phrased in section~\ref{sec:theorem} is discussed first. While in the original Gibbons-Hawking proposal of~\cite{GH} the entropy was extracted from the on-shell action of the sphere saddle of Einstein gravity with a positive cosmological constant, in the present paper, the entropy is extracted from the entire on-shell, effective action of perturbative string theory.  Second, even though no independent definition of the entropy of the de Sitter cosmological horizon exists within the worldsheet formalism, our experience with black holes teaches that $\alpha'$ effects do not nullify an entropy that is nonzero in the supergravity limit $\frac{\alpha'}{r_0^2} \to 0$, where $r_0$ denotes the horizon scale, suggesting that if de Sitter exists in string theory, $\alpha'$ effects do not bring a super-gravitational Gibbons-Hawking entropy down to zero. In particular, the interpretation of that entropy in terms of uncertainty about the region outside the horizon~\cite{GH2} supports the proposition that a nonzero entropy is associated to hypothetical stringy de Sitter horizon in the classical limit $g_s\to0$ and exactly in $\alpha'$.
	
	In relation to possibility (a), if the proposition that the dimension of the Hilbert space of closed cosmologies is one, as suggested in \cite{Marolf:2020xie},\cite{McNamara:2020uza}, it would be consistent that perturbative string theory possesses a direct product of de Sitter and a closed manifold solution without satisfying the version of the Gibbons-Hawking proposal in assumption~\ref{sec:theorem}. As mentioned in the introduction, however, reference~\cite{Dine:1985he} provided an independent argument against the existence of de Sitter in perturbative string theory (more on that below). Another comment concerns the $\text{SU}(2)_k$ WZW model, which describe string propagation in $\mathbb{S}^3$ target spaces when the level is asymptotically large.  This constitutes a counterexample to the conclusion of the argument that a Euclidean de Sitter does not exist. Note that the Lorentzian continuation of the $\text{SU}(2)_k$ WZW model is characterized by an imaginary $H_3$-flux, making a string theory based on a target space of this sort non-unitary. In this example, the version of the Gibbons-Hawking proposal is not satisfied. \footnote{This was pointed out in an earlier paper~\cite{Yoav}.}
		
	An implication of possibility (b) is that if one wants to find a de Sitter solution within string theory, then one should either add boundaries to the compact $(D-d)$ manifold, or rely on non-perturbative corrections in $\alpha'$ and/or $g_s$. Such non-perturbative corrections are tiny provided $g_s \ll 1$ and $\alpha' H^2 \ll 1$, and under these hierarchies, they are not expected to modify the exclusion of the de Sitter solution. 
	
	Another comment is that if one interprets the Dine-Seiberg~\cite{Dine:1985he} result as saying that de Sitter vacua are absent when the string coupling is parametrically small, then the possibility (b) supports this interpretation from a different angle, supposing that the version of the Gibbons-Hawking proposal should be treated as a fundamental law of physics. 
	
	 A different approach is first to note that ``$g_s \ll 1$'' at least in the patch of the Universe in which we live - this is consistent with the fact that quantum-gravitational effects are very difficult to observe, and in fact have never been observed directly. Then one can reason that the argument of this paper supports the fits of DESI~\cite{DESI:2025zgx} that favor deviations from asymptotically de Sitter spacetime, from an independent, theoretical perspective.

\subsection*{Acknowledgements}
Many thanks to Ivano Basile, Miguel Montero, Savdeep Sethi and Arkady Tseytlin, and especially to Eran Palti and Aron Wall for comments that helped improve aspects of the argument. I also thank Ofer Aharony, Ramy Brustein, Sunny Itzhaki, and Thomas Colas for discussions on more general issues.  YZ is supported by the Israel Science Foundation, grant number 1099/24. 
\appendix
\section{Type IIB and 11D supergravities}
\label{app:sug}
\subsection{Type IIB supergravity}
The field content of Type IIB supergravity is the dilaton $\Phi$, metric $G$, NS-NS two-form $B_{2}$, RR potentials $C_{2j}$, dilatino $\psi$ and gravitino $\chi$. 
In the democratic formalism, the bosonic sector of the theory is described by the (pseudo) action
\begin{align}
	\label{IIB}
	 I_{\text{IIB}} = \frac{1}{2\kappa_{10}^2} \int_{M} d^{10}x \sqrt{-G} e^{-2\Phi} \Big( R + 4\partial_{\mu} \Phi \partial^{\mu} \Phi - \frac{1}{2} |H_3|^2\Big)-\frac{1}{8\kappa_{10}^2}\sum_{j=1} ^{5}\int_{M} F_{2j-1} \wedge *F_{2j-1}~, 
\end{align}
where the NS-NS three-form field strength is
\begin{equation}
	H_3 = dB_2~~,
\end{equation}
and the RR field strength forms are given by
\begin{equation}
	\label{F2j}
	F_{2j-1} = dC_{2j-2} -H_3 \wedge C_{2j-4} ~,~  2\leq j \leq 5~,~
\end{equation}
and 
\begin{equation}
	F_1 = dC_0~.
\end{equation}
Additionally, self-duality relations are imposed:
\begin{equation}
	*F_{2j-1} = (-1)^{j-1} F_{11-2j}~,~1\leq j \leq 5~.
\end{equation}
In particular, $F_5$ is self-dual. Bianchi identities are:
\begin{equation}
	dH_3=0 ~,~ dF_1=0~,~ dF_{2j-1} = -H_3 \wedge F_{2j-3}~,~2\leq j \leq 5~~.
\end{equation}
Plugging Eq.~(\ref{F2j}) into Eq.~(\ref{IIB}) gives rise to
\begin{align}
	\label{IIB2}
	I_{\text{IIB}} =& \frac{1}{2\kappa_{10}^2} \int_{M} d^{10}x \sqrt{-G} e^{-2\Phi} \Big( R + 4\partial_{\mu} \Phi \partial^{\mu} \Phi - \frac{1}{2} |H_3|^2\Big)-\frac{1}{8\kappa_{10}^2}\int_{M} dC_0 \wedge *F_1\nonumber\\
	&-\frac{1}{8\kappa_{10}^2}\sum_{j=2} ^{5}\int_{M} (dC_{2j-2}-H_3\wedge C_{2j-4}) \wedge *F_{2j-1}~. 
\end{align}
Integrating by parts the term proportional to $\int_M \sum_{j=1} ^{5} dC_{2j-2}\wedge *F_{2j-1}$ using the graded Leibniz rule produces
\begin{align}
	\label{IIB3}
	I_{\text{IIB}} &= \frac{1}{2\kappa_{10}^2} \int_{M} d^{10}x \sqrt{-G} e^{-2\Phi} \Big( R + 4\partial_{\mu} \Phi \partial^{\mu} \Phi - \frac{1}{2} |H_3|^2\Big)+\frac{1}{8\kappa_{10}^2}\int_{M} C_0 \wedge d*F_1\nonumber\\
	&+\frac{1}{8\kappa_{10}^2}\sum_{j=2} ^{5}\int_{M} \Big( C_{2j-2}\wedge d*F_{2j-1}+ H_3\wedge C_{2j-4} \wedge *F_{2j-1} \Big) - \frac{1}{8\kappa_{10}^2}\sum_{j=1} ^{5}\int_{\partial M} C_{2j-2}\wedge *F_{2j-1}~~. 
\end{align}
Next, the on-shell action is simplified using two types of equations.
First, the dilaton equation is
\begin{align}
	\label{dilaton}
	 R +4\nabla^2 \Phi- 4\partial_{\mu} \Phi \partial^{\mu} \Phi - \frac{1}{2} |H_3|^2=0~.
\end{align}
Second, the RR potential equations are
\begin{equation}
	\label{RR}
	d*F_{2j-1} = -H_3 \wedge * F_{2j+1}~,~1\leq j\leq 4 ~,~ 
\end{equation}
\begin{equation}
	\label{RR2}
	d*F_9=0~~.
\end{equation}
Substituting Eqs.~(\ref{dilaton}), (\ref{RR}) and (\ref{RR2}) into Eq.~(\ref{IIB3}) results in the following on-shell action:
\begin{align}
	\label{IIB4}
	I_{\text{IIB,on-shell}} &= \frac{1}{2\kappa_{10}^2}\int d^{10}x \sqrt{-G} e^{-2\Phi} \Big( 8 \partial_{\mu} \Phi \partial^{\mu} \Phi - 4 \nabla^2 \Phi\Big)-\frac{1}{8\kappa_{10}^2}\sum_{j=1} ^5\int_{\partial M} C_{2j-2}\wedge * F_{2j-1}\nonumber\\
	&+\frac{1}{8\kappa_{10}^2}\int_M \Big( \sum_{j=1}^{4}C_{2j-2}\wedge \big(- H_3 \wedge *F_{2j+1}\big) +\sum_{j=2} ^{5}H_3 \wedge C_{2j-4}\wedge *F_{2j-1}\Big)~.  
\end{align}
By writing $\nabla^2 \Phi = \frac{1}{\sqrt{-G}}\partial_{\mu} \Big( \sqrt{-G} \partial^{\mu} \Phi\Big)$, one can show that the first line in Eq.~(\ref{IIB4}) is a boundary term. Also, in the second line of Eq.~(\ref{IIB4}), one can shift the index of the second sum on the right by writing $j=k+1$. These steps yield
\begin{align}
	\label{IIB5}
	I_{\text{IIB,on-shell}} &= \frac{1}{\kappa_{10}^2}\int_{\partial M} \Big[ d^{9}x \sqrt{-h} n^{\mu} \partial_{\mu} \big(e^{-2\Phi}\big) -\frac{1}{8}\sum_{j=1} ^5 C_{2j-2}\wedge * F_{2j-1}\Big]\nonumber\\
	&+\frac{1}{8\kappa_{10}^2}\int_M \Big(- \sum_{j=1}^{4}C_{2j-2}\wedge  H_3 \wedge *F_{2j+1} +\sum_{k=1} ^{4}H_3 \wedge C_{2k-2}\wedge *F_{2k+1}\Big)~.  
\end{align}
The notation $h$ is the determinant of the induced metric on the boundary of the target spacetime.
Clearly, the last line of Eq.~(\ref{IIB5}) vanishes on the ground that $H_3 \wedge C_{2j-2} = C_{2j-2} \wedge H_3$, as $C_{2j-2}$ is a tensor of even rank. The result is that the on-shell Type IIB supergravity action is a surface term:
\begin{align}
	\label{IIB6}
	I_{\text{IIB,on-shell}} &= \frac{1}{\kappa_{10}^2}\int_{\partial M} \Big[ d^{9}x \sqrt{-h} n^{\mu} \partial_{\mu}\big( e^{-2\Phi}\big) -\frac{1}{4}\sum_{j=1} ^5 C_{2j-2}\wedge * F_{2j-1}\Big]~.  
\end{align}
If the gauge potentials are globally well-defined, this action is gauge-invariant for gauge parameter forms that vanish on the boundary $\partial M$ (in the absence of ``edge modes''). If, on the other hand, the gauge potentials are not globally well-defined, one should partition the manifold $M$ into patches and evaluate the boundary term as a sum of individual boundary components whose union is $\partial M$. To make the variational principle well-defined, one adds a Gibbons-Hawking-York boundary term. 

When $F_1=0,~F_3=0$, the self-duality of $F_5$ implies that $F_5 \wedge F_5=0$ and the Type IIB action reduces to a boundary term that vanishes for constant dilaton backgrounds - see~\cite{Kurlyand:2022vzv}. To summarize, Eq.~(\ref{IIB6}) expresses the on-shell action of Type IIB supergravity in the democratic formalism as a surface term. 
\subsection{11D supergravity}
The field content of 11D supergravity is the metric $G$, the three-form potential $A_3$, and the gravitino field $\chi$. The bosonic action of the theory is
\begin{align}
	\label{11D}
	I_{11D} =\frac{1}{2\kappa_{11}^2}\int_M d^{11}x \sqrt{-G} \Big(R -\frac{1}{2} |F_4|^2\Big)-\frac{1}{12\kappa_{11}^2}\int_M A_3 \wedge F_4 \wedge F_4~,
\end{align}
where the field strength of $A_3$ is defined in
\begin{equation}
	F_4 \equiv dA_3~.
\end{equation}
The Einstein field equations are
\begin{align}
	\label{Ein}
	R_{\mu \nu} - \frac{1}{2} G_{\mu \nu} R = \frac{1}{12} F_{\mu \alpha \beta \gamma} F_{\nu} ^{~~\alpha \beta \gamma} -\frac{1}{96} G_{\mu \nu} F_{\alpha \beta \gamma \delta} F^{\alpha \beta \gamma \delta}.
\end{align}
The three-form potential equations are
\begin{equation}
	\label{F4} 
	d*F_4 = -\frac{1}{2} F_4 \wedge F_4~.
\end{equation}
The Bianchi identity is 
\begin{equation}
	dF_4 =0~.
\end{equation}
The trace of Eq.~(\ref{Ein}) implies
\begin{align}
	\label{Ricc}
	 R = \frac{1}{6}|F_4|^2~.
\end{align}
Substituting Eqs.~(\ref{Ricc}) and (\ref{F4}) into Eq.~(\ref{11D}) yields
\begin{align}
	\label{11Db} 
	I_{11D,\text{on-shell}}=-\frac{1}{6\kappa_{11}^2}\int_M d^{11} x \sqrt{-G} |F_4|^2 +\frac{1}{6\kappa_{11}^2}\int_M A_3 \wedge d*F_4~.
\end{align}
Now, since $A_3$ is an odd-form field, the graded Liebniz rule implies
\begin{equation}
	d(A_3 \wedge * F_4) = F_4 \wedge * F_4 - A_3 \wedge d*F_4~.
\end{equation}
Hence, Eq.~(\ref{11Db}) becomes
\begin{align}
	\label{11Dc} 
	I_{11D,\text{on-shell}}=-\frac{1}{6\kappa_{11}^2}\int_{\partial M}  A_3 \wedge *F_4 ~.
\end{align}
Typically, one adds another boundary term, the Gibbons-Hawking-York term, to ensure that the variational principle is well defined, which entails that when imposing that metric variations vanish on $\partial M$, the variation of the action vanishes when the equations of motion are satisfied~\cite{GH}. To compare with Eq.~(\ref{IIB6}) from the previous subsection, though it came from a different theory, the Gibbons-Hawking-York term leads to a term that goes like $\sqrt{-h} n^{\mu}\partial_{\mu}\big( e^{-2\Phi}\big)$ in the action, after performing a dimensional reduction on the eleventh dimension. The conclusion is that, also in 11D supergravity, the on-shell action is a boundary term. Thus, one can run the argument in section~\ref{sec:proof} in the context of this classical theory. 
\section{Quantum effective action approach}
\label{app:Ieff}
This Appendix considers an alternative approach to the string loop expansion, wherein perturbative $g_s$ effects are incorporated into the following action 
	\begin{equation}
	\label{I3}
	I= \int_M d^{D} x ~\Bigg(e^{-2\Phi} L_{-2}+\sum_{g=0} ^{\infty} e^{g\Phi} L_g\Bigg)~~.
\end{equation}
The same action was written in reference~\cite{Tseytlin:1988tv} Eq.~(27) and is in general non-local, and the sum over $g$ does not converge. The non-locality will not affect the conclusion, and the absence of convergence would be fixed if one were to incorporate non-perturbative effects, though they are suppressed in the present discussion.  

The dilaton equation obtained from varying action~(\ref{I3}) is given by
\begin{align}
	\label{dil2} 
	-2 e^{-2\Phi} L_{-2}+\sum_{g=0} ^{\infty} ge^{g\Phi} L_{g} &=\partial_{\mu} \Big(e^{-2\Phi}\frac{\delta L_{-2}}{\delta (\partial_{\mu} \Phi)} \Big)-\partial_{\mu} \partial_{\nu}\Big(e^{-2\Phi}\frac{\delta L_{-2}}{\delta (\partial_{\mu} \partial_{\nu} \Phi)} \Big)+... \nonumber\\
	& +\sum_{g=0} ^{\infty}\Bigg[\partial_{\mu} \Big( e^{g\Phi}\Big(\frac{\delta L_g}{\delta (\partial_{\mu} \Phi)} \Big)\Big)-\partial_{\mu} \partial_{\nu}\Big(e^{g\Phi}\frac{\delta L_{g}}{\delta (\partial_{\mu} \partial_{\nu} \Phi)} \Big)+...\Bigg]~.
\end{align}
Utilizing Eq.~(\ref{dil2}) to express $ e^{-2\Phi}L_{-2}$ in terms of other terms, the resulting on-shell action is 
\begin{align}
	\label{Icl2}
	I_{\text{on-shell}} &=\int_M d^D x\sum_{g=0} ^{\infty} \Big(1+\frac{1}{2}g\Big)e^{g\Phi} L_{g} -\frac{1}{2} \int_M d^{D} x ~\partial_{\mu} \Bigg(e^{-2\Phi}\frac{\delta L_{-2}}{\delta (\partial_{\mu} \Phi)}-\partial_{\nu}\Big(e^{-2\Phi}\frac{\delta L_{-2}}{\delta (\partial_{\mu} \partial_{\nu} \Phi)}\Big)+...\Bigg)\nonumber\\
	&~~~-\frac{1}{2}\int_M d^D x\sum_{g=0} ^{\infty}\Bigg[\partial_{\mu}\Big(   e^{g\Phi}\frac{\delta L_g}{\delta (\partial_{\mu} \Phi)} \Big)-\partial_{\mu} \partial_{\nu}\Big(e^{g\Phi}\frac{\delta L_{g}}{\delta (\partial_{\mu} \partial_{\nu} \Phi)} \Big)+...\Bigg]\nonumber\\
 &= \int_M d^D x\sum_{g=0} ^{\infty} \Big(1+\frac{1}{2}g\Big)e^{g\Phi} L_{g}-\frac{1}{2} \int_{\partial M} d^{D-1} x ~n_{\mu} \Bigg(e^{-2\Phi}\frac{\delta L_{-2}}{\delta (\partial_{\mu} \Phi)}-\partial_{\nu}\Big(e^{-2\Phi}\frac{\delta L_{-2}}{\delta (\partial_{\mu} \partial_{\nu} \Phi)}\Big)+...\Bigg)
	\nonumber\\
	& -\frac{1}{2}\int_{\partial M} d^{D-1} x~n_{\mu}\sum_{g=0} ^{\infty}\Bigg[  e^{g\Phi}\frac{\delta L_g}{\delta (\partial_{\mu} \Phi)} - \partial_{\nu}\Big(e^{g\Phi}\frac{\delta L_{g}}{\delta (\partial_{\mu} \partial_{\nu} \Phi)} \Big)+...\Bigg]~~.
\end{align}
Therefore, the action $I_{\text{on-shell}}$ is a sum of perturbative terms in $g$ starting with $g=0$, and boundary terms.  In the special case where $M$ is a closed Euclidean space, and invoking smoothness and the well-defined nature of the integrands involved, the surface terms drop, and it follows that
\begin{equation}
	\label{I03}
	I_{\text{on-shell}}=\int_M d^D x\sum_{g=0} ^{\infty} \Big(1+\frac{1}{2}g\Big)e^{g\Phi} L_{g}~.
\end{equation}

\end{document}